\documentclass[12pt]{article}
 \usepackage{graphicx}

\usepackage{scicite}
   
\usepackage{times}

\newenvironment{sciabstract}{%
\begin{quote} \bf}
{\end{quote}}

\newcounter{lastnote}
\newenvironment{scilastnote}{%
\setcounter{lastnote}{\value{enumiv}}%
\addtocounter{lastnote}{+1}%
\begin{list}%
{\arabic{lastnote}.}
{\setlength{\leftmargin}{.22in}}
{\setlength{\labelsep}{.5em}}}
{\end{list}}

\title{Inter--areal coordination of \\ 
columnar architectures during \\
visual cortical development}

\author
{Matthias Kaschube,$^{1,2}$ Michael Schnabel,$^{1}$ 
Siegrid L\"owel,$^{3}$ \\ Fred Wolf$^{1\ast}$ \\\\
\normalsize{$^{1}$Max-Planck-Institute for Dynamics and
  Self-Organization, and}\\ 
\normalsize{Bernstein-Center for Computational Neuroscience,}\\
\normalsize{Bunsenstrasse 10, 37073, G\"ottingen, Germany.}\\ 
\normalsize{$^{2}$Lewis-Sigler Institute for Integrative Genomics, and Physics Department,}\\
\normalsize{ Princeton University, 261 Carl-Icahn Laboratory, Princeton NJ, 08544, USA}\\
\normalsize{$^{3}$Institute of General Zoology and Animal Physiology,
  Friedrich-Schiller-University}\\ 
\normalsize{Erbertstrasse. 1, 07743 Jena, Germany.}\\ 
\normalsize{$^\ast$To whom correspondence should be addressed; E-mail: fred@nld.ds.mpg.de}
}

\date{}

\begin{document} 


\baselineskip24pt


\maketitle


\begin{sciabstract}

The occurrence of a critical period of plasticity in the visual cortex
has long been established, yet its function in normal development is not fully understood. Here we show that as the late phase of the critical period unfolds,
different areas of cat visual cortex develop in a coordinated manner.
Orientation columns in areas V1 and V2 become matched in size in
regions that are mutually connected. The same age trend is found for such
regions in the left and right brain hemisphere. Our results indicate that a function of critical period plasticity is to progressively coordinate the functional architectures of different cortical areas - even across hemispheres.
\end{sciabstract}

\pagebreak


Sensory input has the capacity to shape neuronal circuits, especially
in the juvenile brain. In the visual cortex, the period when circuits 
are particularly susceptible to changes of visual input typically
lasts many weeks as highlighted by the critical period for the effects
of monocular
deprivation\cite{blakemore:74,olson:80,jones:84}. The onset of this period is
delayed relative to the onset of visual
experience\cite{crair:01,sur:01,katz:02} and
 key response properties of neurons
are already present in almost adult form well before critical period onset\cite{crair:98,crowley:00}. 
What then is the function of a period of relatively strong
plasticity at such a late stage of development?
Evidently, it enables the cortex to adapt to deprivations
from visual input\cite{olson:80,jones:84,sengpiel:99} that may be
caused by injuries or disease or,  
under normal conditions, by the shadows of retinal blood
vessels\cite{adams:02}. By 
the same token, however, short periods 
of abnormal vision can cause permanent deficits. The
paradox of the critical
period of plasticity is, thus, that it seems to provide only little benefit compared
to its great potential for handicap\cite{pettigrew:78}. Alternatively,
it has often been suggested 
that a period of plasticity plays an important role for the development
of normal vision by refining neural circuits through structured patterns
of cortical activity. In fact, under normal
experience, vision improves during this period\cite{giffin:78}, but
no evidence for a  
reorganization  of visual cortical representations has been
observed so far\cite{chapman:96,sengpiel:98}.

Here, we show that in cat visual cortex columnar architectures of different cortical areas
develop in a coordinated manner as the late phase of the period of
plasticity unfolds. Orientation columns analyzing contours in different
parts of the visual field constitute a repetitive structure in visual
cortical areas V1 and V2\cite{hubel/wiesel:62,levay/nelson:91} (Fig.
1, A and B). Orientation columns in animals aged between 6 and 15 weeks were labeled with 2-{[}$^{14}$C{]}-deoxyglucose 
(2-DG)\cite{loewel:02} and visualized in flat-mount sections of cat
visual cortical areas V1 and V2
\cite{loewel:87,loewel/etal:88,loewel/singer:90} 
(n=41 brain hemispheres; N=27 animals). Both V1 and V2 contain a complete
topographic representation of the contralateral visual
field\cite{tusa:78,tusa:79}. 
Columns at topographically corresponding locations in the two areas
represent similar visual field positions and are selectively and
mutually connected by corticocortical connections\cite{salin:95}.
This topography of the two areas enables us to conveniently compare
the layout of distant columns that are mutually connected and represent
related aspects of the sensory input. We analyzed the spacing of orientation
columns locally using a recently developed wavelet method\cite{kaschube:02}.
Because this method provides highly precise estimates of local column
spacing with an error much smaller than the large intrinsic variability\cite{shmuel:00,kaschube:02}
of column spacings (SEM, 15--50$\mu$m), differences and similarities
of column spacings in the sample can be identified reliably. 

We first analyzed the mean spacing of orientation columns in areas
V1 and V2 and assessed their statistical and age dependence. We found that
mean column spacings, $\Lambda$, varied considerably in different individuals
(Fig. 1C). In V1 values ranged between 1.1mm and 1.4mm, in V2 between
1.2mm and 1.8mm. The distributions for the two areas were partially
overlapping with the smallest column spacings from V2 at about the
average value of V1. Nevertheless, in all hemispheres, the mean column
spacing, $\Lambda$, in V2 was substantially larger than in V1, consistent
with previous reports\cite{loewel:87}. Mean column spacings, $\Lambda$,
did not vary independently across different animals in V1 and V2,
but were substantially correlated in both areas (Fig 1D; r=0.62, p$<10^{-5}$,
permutation test). A dependence on the age of animals, however, was
not observed, neither for mean column spacings, $\Lambda$, (see also
Fig. 5F) nor for their differences between V1 and V2 (data not shown).

An age dependence of column spacings became apparent, when we decomposed
  the spatial variation and covariation of column
  spacings in areas V1 and V2 into different components. 
In both areas V1 and V2, orientation columns generally exhibited a
substantial intra-areal variation in spacing around the mean column
spacing. One part of this variation is common to all
hemispheres. In the following, we will call this the systematic
topographic component of column spacings (the blue map in the supplementary
Fig. S1). It is the intra-areal variation (in V1 or V2) averaged over
the entire population of hemispheres, thus it is a 2D spacing map with zero
mean. For averaging, the V1/V2 borders of different hemispheres were
aligned and maps from right hemispheres were mirror-inverted. The
remaining component of variation characterizes an individual hemisphere
and is called individual topographic component 
of column spacings in the following  (the orange map in Fig. S1). It is also a 2D spacing map with zero mean calculated by subtracting 
from the map of local column spacing of an area its mean and the systematic component. The
variances of the systematic component, of the individual topographic component, and of the mean column spacing add up to the total
variance of local column spacings in the sample. The
individual topographic component accounted for the largest part of the
variance of column spacings in both areas V1 and V2 (Supplementary
Fig. S1D). As shown below examining this bigest variance component reveals that column sizes at topographically corresponding locations in different areas become better matched as the critical period progresses. First, however, we will examine the systematic component which demonstrates that orientation columns in the two areas are coordinated on average.

In V1, the systematic component exhibited 
virtually the same overall 2D organization as the one in V2, appearing
as a horizontally stretched mirror-image of the V2 map when displayed
side by side (Fig. 2, A and B). The systematic variation in areas
V1 and V2 ranged between -0.15mm and +0.15mm (Fig. 2, A and B). In
both areas, columns were systematically wider than average along the
representation of the horizontal meridian (HM) with this tendency
increasing towards the periphery. In contrast, columns smaller than
average prevailed along the peripheral representations of
the vertical meridian (VM). In order to conveniently compare topographically
corresponding parts in areas V1 and V2, the V2 map was mirror-inverted
and morphed by superimposing major landmarks such as the representations
of the VM (located along the V1/V2 border), the central visual field,
and the HM. The morphed V2 map strongly resembled the V1 map
(compare Fig. 2, B and C), and the cross-correlation between the maps was
high (r=0.66). Furthermore, the systematic component observed in a
comparable data set of ocular dominance column spacings in cat V1
(Fig. 2D, modified from\cite{kaschube:03}) also exhibited a very
similar intra-areal organization with a strong cross-correlation of
r=0.82 to the population averaged spacing map for orientation columns
in V1 (compare Fig. 2, B and D). No significant age dependence was
  found for the systematic component of column spacings when
  calculated separately for groups of younger and older animals.

Next, we analyzed the individual topographic component of local column
spacings.
Like the systematic component, the individual topographic component was
often similar  in V1 and V2 in regions  analyzing the same part of 
the visual field and being mutually connected. The examples shown in
Fig. 3, A to C, display the 
same general pattern in both areas V1 and V2 with maxima (white) and
minima (dark orange) at approximately  corresponding retinotopic locations.
More examples are shown in Supplementary Fig. S2. Among different
brains the 
pattern of individual component differed considerably. To quantify
the similarity of the individual topographic component in V1 and V2
we calculated, for each hemisphere, the absolute value of the difference
between both maps averaged over all analyzed locations, called their
mismatch $\Delta_{{V1V2}}$. These mismatches, $\Delta_{{V1V2}}$, were significantly
smaller than values obtained for randomly assigned pseudo V1/V2 pairs
(p=0.03, permutation test). Thus, in an individual hemisphere the
individual topographic component is coordinated at
topographically corresponding locations of both areas.

Intriguingly, this column size matching also applied to columns at corresponding locations in left right pairs of areas from both hemispheres.
Whereas maps of the individual topographic component in pairs of both hemispheres
differed at mirror-symmetric locations, they were often very similar
along the V1/V2 border, i.e. in the region containing the cortical representations of the VM (Fig. 3E). V1-columns of representations of the VM in both hemispheres receive similar afferent input and also mutual input mediated by callosal connections concentrated in the vicinity of the VM representation\cite{olavarria:01,olavarria:02}.
By the mismatch, $\Delta_{LR}$, we quantified the absolute value of differences between left and right maps averaged over a strip of 3mm width adjacent to the V1/V2 border. Values of $\Delta_{LR}$ were
significantly smaller than those obtained for randomly assigned pairs
of hemispheres, (p=0.01, permutation test). A similar behavior was
found for V2 (data not shown). Missmatches $\Delta_{LR}$ were
larger for regions more distant from the V1/V2 border. 

Analyzing the dependency of column size matching on age, we found that 
it improves between different areas during the late phase of the critical
period. Examples are shown in Fig. 4, A and B.  Whereas in younger 
animals the pattern of individual topographic component  differed in V1
and V2 (Fig. 4A) and along the V1/V2 border (Fig. 4B), it was relatively
similar in older animals.  Fig. 4, C to E, shows 
the column spacing mismatches $\Delta$ 
for topographically corresponding columns in V1/V2 pairs and in
left/right pairs of V1 and of V2 as a function of age. For all three pairs of
areas substantial mismatches were only observed in animals younger
than 10 weeks. In older animals mismatches of column spacing were
in general less than 0.1mm. Consequentially, all three measures were
significantly anti-correlated with animal age ($\Delta_{V1V2}$,
$r=-0.39$, $p=0.01$; $\Delta_{LR}$ for 
V1, $r=-0.64$, $p=0.007$; for V2, $r=-0.50$, $p=0.02$). In contrast,
the average column spacing in areas V1 and V2 was independent of age
(Fig. 4F). Thus, whereas the average column spacing in areas V1 and V2 remained constant, locally, the column spacing increased or decreased such that
mutually connected columns in different areas became coordinated in size.

Previous studies emphasized the apparent
stability of the columnar architecture during early visual development\cite{chapman:96,sengpiel:98}. 
Unlike these studies, but fully consistent with them, we observe changes of the columnar architecture during the late phase of the period 
of visual cortical plasticity. Assessed by  monocular deprivation, the
critical period  peaks at postnatal week 6 and slowly decreases
afterward\cite{olson:80,jones:84}. In addition, both the 
connections from V1 to V2 and the callosal connections undergo a
process of refinement over this age range. Densities of connections are maximal between week 4 and week 10 and then gradually decline over the following months\cite{price:94,innocenti:96}. We observed changes in three pairs of visual areas, in the two areas V1 from the left and
right brain hemispheres, in the two areas V2, and in areas V1 and V2
from the same hemisphere. The changes involve large
parts of each of these areas and result in refined coordination of column sizes among mutually connected regions. 

The trend to minimize
the mismatch of column sizes between different areas suggests a process
of optimization of columnar architectures that is not confined to individual areas, but potentially spans the entire visual system. Known  mechanism may underlay the emergence of coordinated
column layouts in widely distributed cortical regions. 
Recently, it has been shown that pharmacologically 
shifting the balance of inhibition and excitation during development
modifies column spacing in cat V1\cite{hensch:04}. Presumably, column
spacings are also determined by the local inhibitory-excitatory balance
in normal development. Here this balance may be set by regulatory
mechanisms such as synaptic scaling\cite{desai:02}, that
are sensitive to neuronal activity.  The emergence of size matched
columns in distant cortical regions might thus be caused by a similar
balance of inhibition and excitation in these regions that emerges
from their mutual synaptic coupling. Because cortical processing in general takes place in networks distributed across many areas, it is conceivable that developing matching of local circuits serving different submodalities is a general characteristic of cortical network formation.

\bibliography{paper9}

\begin{thebibliography}{10}

\bibitem{blakemore:74}
C.~Blakemore, R.~V. Sluyters, {\it J. Physiol.\/} {\bf 237}, 195 (1974).

\bibitem{olson:80}
C.~Olson, R.~Freeman, {\it Exp Brain Res\/} {\bf 39}, 17 (1980).

\bibitem{jones:84}
K.~Jones, P.~Spear, L.~Tong, {\it J Neurosci\/} {\bf 4}, 2543 (1984).

\bibitem{crair:01}
M.~Crair, J.~Horton, A.~Antonini, M.~Stryker, {\it J. Comp. Neurol.\/} {\bf
  430}, 235 (2001).

\bibitem{sur:01}
M.~Sur, C.~Leamey, {\it Nat. Rev. Neurosci.\/} {\bf 2}, 251 (2001).

\bibitem{katz:02}
L.~Katz, J.~Crowley, {\it Nat. Rev. Neurosci.\/} {\bf 3}, 34 (2002).

\bibitem{crair:98}
M.~Crair, D.~Gillespie, M.~Stryker, {\it Science\/} {\bf 279}, 566 (1998).

\bibitem{crowley:00}
J.~Crowley, L.~Katz, {\it Science\/} {\bf 290}, 1321 (2000).

\bibitem{sengpiel:99}
F.~Sengpiel, P.~Stawinski, T.~Bonhoeffer, {\it Nature Neuroscience\/} {\bf 2},
  727 (1999).

\bibitem{adams:02}
D.~Adams, J.~Horton, {\it Science\/} {\bf 298}, 572 (2002).

\bibitem{pettigrew:78}
J.~D. Pettigrew, {\it Neuronal Plasticity\/}, C.~W. Cotman, ed. (Raven Press,
  NY, 1978), pp. 311--330.

\bibitem{giffin:78}
F.~Giffin, D.~Mitchell, {\it J. Physiol.\/} {\bf 274}, 511 (1978).

\bibitem{chapman:96}
B.~Chapman, M.~P. Stryker, T.~Bonhoeffer, {\it J. Neurosci.\/} {\bf 16}, 6443
  (1996).

\bibitem{sengpiel:98}
F.~Sengpiel, {\it et~al.\/}, {\it Neuropharmacology\/} {\bf 37}, 607 (1998).

\bibitem{hubel/wiesel:62}
D.~H. Hubel, T.~N. Wiesel, {\it J. Physiol.\/} {\bf 160}, 215 (1962).

\bibitem{levay/nelson:91}
S.~LeVay, S.~B. Nelson, {\it Vision and Visual Dysfunction\/} (Macmillan,
  Houndsmill, 1991), chap.~11, pp. 266--315.

\bibitem{loewel:02}
S.~L\"owel, {\it The Cat Primary Visual Cortex\/}, B.~Payne, A.~Peters, eds.
  (Academic Press, San Diego, 2002), chap.~2, pp. 167--193.

\bibitem{loewel:87}
S.~L\"owel, B.~Freeman, B.~Singer, {\it J. Comp. Neurol.\/} {\bf 255}, 401
  (1987).

\bibitem{loewel/etal:88}
S.~L\"owel, H.-J. Bischof, B.~Leutenecker, W.~Singer, {\it Exp. Brain Res.\/}
  {\bf 71}, 33 (1988).

\bibitem{loewel/singer:90}
S.~L\"owel, W.~Singer, {\it Dev. Brain Res.\/} {\bf 56}, 99 (1990).

\bibitem{tusa:78}
R.~Tusa, L.~Palmer, A.~Rosenquist, {\it J. Comp. Neurol.\/} {\bf 177}, 213
  (1978).

\bibitem{tusa:79}
R.~Tusa, A.~Rosenquist, L.~Palmer, {\it J. Comp. Neurol.\/} {\bf 185}, 657
  (1979).

\bibitem{salin:95}
P.~Salin, H.~Kennedy, J.~Bullier, {\it Can. J. Physiol. Pharmacol.\/} {\bf 73},
  1339 (1995).

\bibitem{kaschube:02}
M.~Kaschube, F.~Wolf, T.~Geisel, S.~L\"owel, {\it J. Neurosci.\/} {\bf 22},
  7206 (2002).

\bibitem{shmuel:00}
A.~Shmuel, A.~Grinvald, {\it Proc. Natl. Acad. Sci.\/} {\bf 97}, 5568 (2000).

\bibitem{kaschube:03}
M.~Kaschube, {\it et~al.\/}, {\it Eur. J. Neurosci.\/} {\bf 18}, 3251 (2003).

\bibitem{olavarria:01}
J.~Olavarria, {\it J. Comp. Neurol.\/} {\bf 433}, 441 (2001).

\bibitem{olavarria:02}
J.~Olavarria, {\it The Cat Primary Visual Cortex\/}, B.~Payne, A.~Peters, eds.
  (Academic Press, San Diego, 2002), chap.~6, pp. 259--318.

\bibitem{price:94}
D.~Price, J.~Ferrer, C.~Blakemore, N.~Kato, {\it J Neurosci\/} {\bf 14}, 2747
  (1994).

\bibitem{innocenti:96}
D.~Aggoun-Aouaoui, D.~Kiper, G.~Innocenti, {\it Eur J Neurosci\/} {\bf 8}, 1132
  (1996).

\bibitem{hensch:04}
T.~Hensch, M.~Stryker, {\it Science\/} {\bf 303}, 1678 (2004).

\bibitem{desai:02}
N.~S. Desai, R.~H. Cudmore, S.~B. Nelson, G.~G. Turrigiano, {\it Nat
  Neurosci.\/} {\bf 5}, 783 (2002).

\end{thebibliography}

\bibliographystyle{Science}


\begin{scilastnote}
\item We thank M. Puhlmann and S. Bachmann for excellent technical
assistance; U. Ernst for providing the morphing program; W. Singer for
his permission to use the 2-DG autoradiographs obtained by S.L. in his
laboratory for quantitative analysis; T. Geisel for fruitful discussions; M. Brecht, D. Brockmann and S. Palmer for helpful comments on an earlier version of this manuscript. Supported by
MPG, HSFP.
\end{scilastnote}

\noindent {\bf Supporting Material}\\
Materials and Methods\\
References\\
Fig. S1, S2\\
Table S1\\


\clearpage

\noindent {\bf Fig. 1.} 
Mean column spacings in V1 and V2 covary.  
({\bf A}) and ({\bf B}) Overall layout of orientation columns in 
V1 and V2 in two individuals (white dashed line: V1/V2 border; scale
bar, 10mm).    
For {\bf A} the mean column spacing $\Lambda$ is
relatively large in both areas  (V1, 1.21mm; V2, 1.58mm), whereas for 
{\bf B} it is small in both areas (V1, 1.09mm; V2, 1.31mm). 
({\bf C}) Mean column spacings $\Lambda$ in V1 (crosses) and V2
(boxes) from n=41 hemispheres (N=27 animals).  
Values in V1 and V2 vary considerably in different hemispheres (V1,
1.0--1.4mm; V2, 1.2--1.8mm).  
({\bf D}) Mean column spacings $\Lambda$ in V1 and V2 are strongly correlated 
(r=0.62, p$<10^{-5}$, permutation test).

\noindent {\bf Fig. 2.} 
The systematic topographic component of column spacing is similar in 
subregions encoding the same visual field position in areas V1 and
V2.
({\bf A} and {\bf B}) Systematic topographic component of
orientation column spacing 
in V2 (A) and V1 (B) (color scale codes systematic
deviation from the
mean value; arrangement and symbols as in Fig.
1). ({\bf  C}) The morphed map from V2 (A).   
({\bf D}) Population averaged spacing of ocular dominance 
columns in cat V1 (modified from \cite{kaschube:03}.
SD for V2, 0.052mm; V1,  0.047mm;  V1 ocular dominance columns, 
0.049mm.  Scale bar, 10mm.
Note that for both orientation and ocular dominance maps columns
representing the HM 
and, in particular, the horizontal periphery were on
average wider  than columns representing the peripheral 
parts of the VM.
Hence, the systematic variations of orientation columns in V1 and V2 were
correlated at topographically corresponding locations (correlation between
(B) and (C), r=0.69), as were those of orientation and
ocular dominance columns in V1 (correlation between (B) and (D), r=0.82).

\noindent {\bf Fig. 3.} 
In individual brains, columns in different areas are closely matched in 
size at topographically corresponding subregions. 
({\bf A}-{\bf C}) Similarity of the individual topographic
component of column spacings  in  V1 and V2. 
({\bf A})
The overall layout of orientation columns for the hemisphere shown in Fig.
1. A pair of topographically corresponding  subregions from the more
anterior part of V1 and V2 (yellow boxes) and
a pair from the more posterior part (blue
boxes) are displayed magnified such that all differences but the individual
variation were equalized.
The relative difference of mean column
spacings in V1 and V2 was $\Lambda_{V2}/\Lambda_{V1}=1.3$. To equalize
this difference, the subregions from V1 were magnified relative to those from
V2 by this factor. To equalize the differences due to the systematic
topographic component the two posterior sub-regions were magnified by an
additional factor of $1.05$.
Note that the spacing of columns is similar within each pair.
({\bf B}) Patterns of individual variation of column spacing for
V2, V1, and the morphed version of V2 for the hemisphere in (A)
(color scale, black cross and contour lines as in Fig. 1).
({\bf C}) Similarity of the individual variation in V1 and V2 at
topographically
matched  locations in another example.  
({\bf D} and {\bf E}) Similarity of the individual topographic
component of column spacings  in the left and right brain hemisphere.
({\bf D}) The overall layout of orientation columns in the left and right
hemisphere of an individual animal. A pair of topographically corresponding
regions  from the anterior part of the VM representation
in V1 of both hemispheres (yellow boxes)
and a pair from a more posterior part in V1 (blue boxes)  was magnified
such that all differences except the individual topographic component
were equalized.
Mean column spacings $\Lambda_{V1}^r$ and $\Lambda_{V1}^l$ were equal
in both hemispheres and the systematic topographic components were equalized
by magnifying the two posterior sub-regions by a factor
$0.98$.
Note that the spacing of columns is  similar within each pair.
({\bf E}) Patterns of the individual variation of local column spacing
in V1 for the hemispheres in (D)
displayed with the representation
of the  VM  side by  side (crosses and contour lines as
in Fig. 1).  In both residual maps,
the blue rectangles (width, 3mm) are positioned at the
representation of the VM.
Note that the residual maps tend to be similar only along the
VM-representations. Scale bar, 10mm.

\noindent {\bf Fig. 4.} 
Consolidation of column size matching with age. 
({\bf A} and {\bf B}) Comparison of individual topographic component of column spacing
in V1/V2 pairs (A) and in left/right pairs from V1 (B) at earlyer and
later ages.
 ({\bf C}) Distances $\Delta_{{V1V2}}$ between the individual 
topographic components  in
V1 and V2 versus age.
({\bf D}) Distances $\Delta_{LR}$ between the V1 individual 
topographic components in the left and right brain hemispheres 
versus age (calculated within the blue 
rectangles in B). 
({\bf E}) Distances $\Delta_{LR}$ for V2 (calculated in the blue
rectangles in B from the morphed maps of V2). 
({\bf F}) Mean column spacings $\Lambda$ in V1 (crosses) and V2
(boxes) (from Fig. 1C) versus age. 
Correlations with age are significant in (C) ($r$=-0.64, p=0.007), 
(D) ($r$=-0.5, p=0.02) and (E) ($r$=-0.39, p=0.01), 
but were not significant in (F) (p$\>$0.05).

\clearpage

\begin{figure}
\begin{center}
\hspace*{-2cm}
\includegraphics[width=6.0cm]{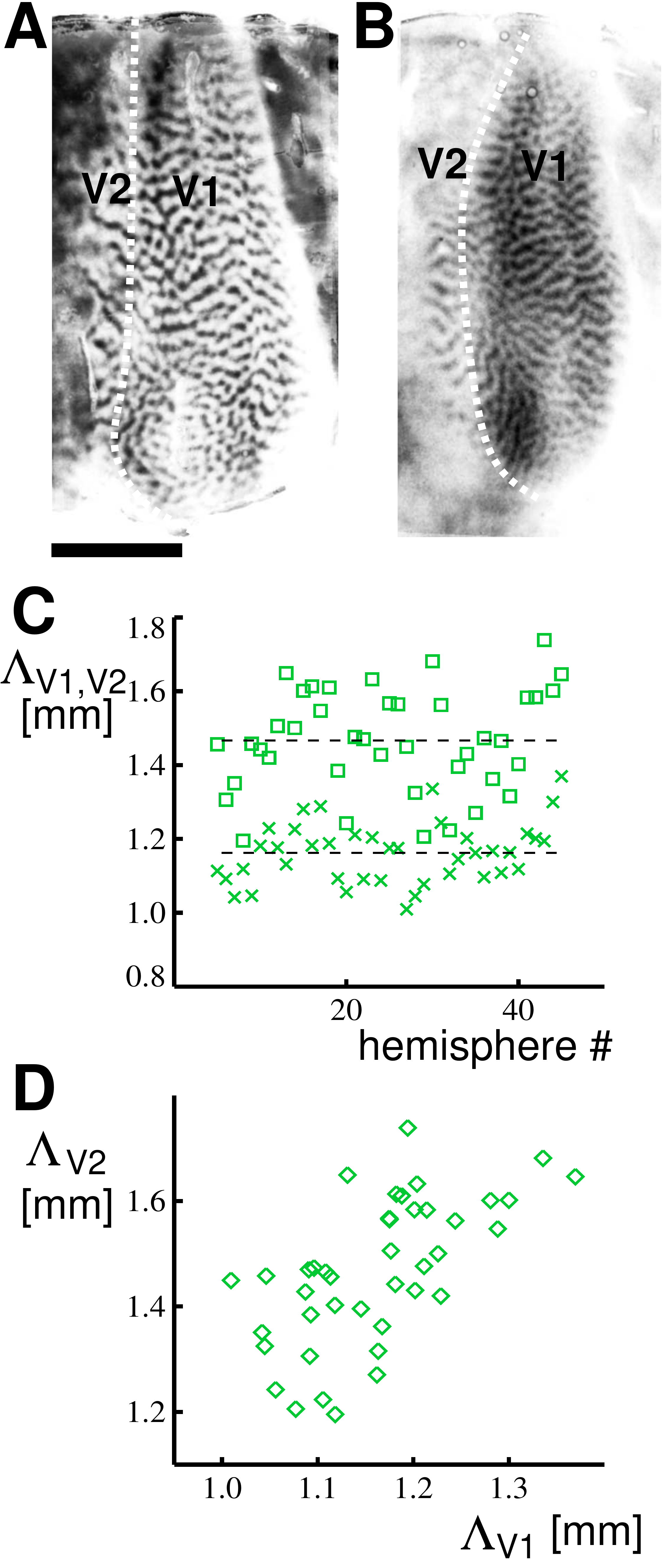}
\vspace{0cm}
\end{center}
 \caption{ }
\end{figure}

\newpage
\begin{figure}
\begin{center}
\includegraphics[width=10cm]{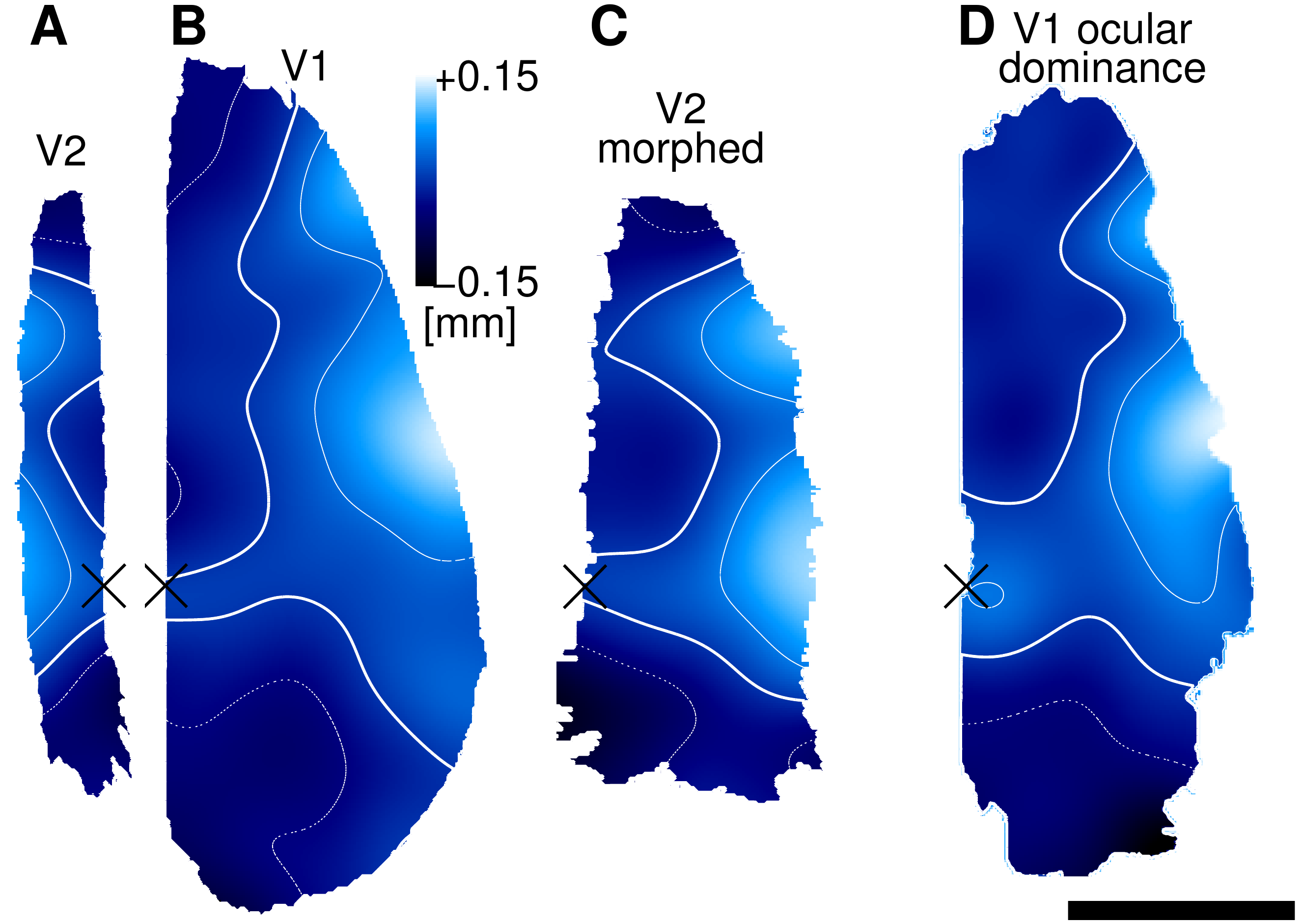}
\vspace{2cm}
\end{center}
 \caption{ }
\end{figure}

\newpage
\begin{figure}
\begin{center}
\includegraphics[bb=0 0 583 390,width=14cm]{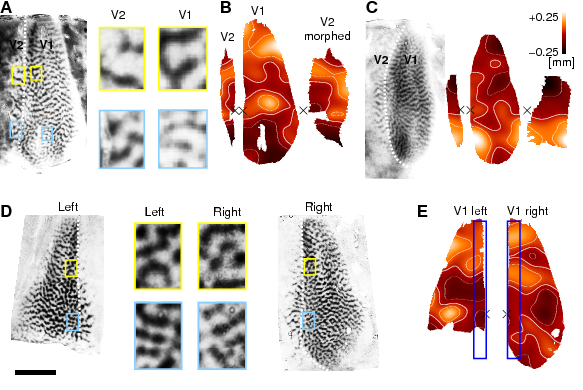}
\vspace{2cm}
\end{center}
 \caption{ }
\end{figure}

\newpage
\begin{figure}
\begin{center}
\includegraphics[bb=0 0 458 569,width=12cm]{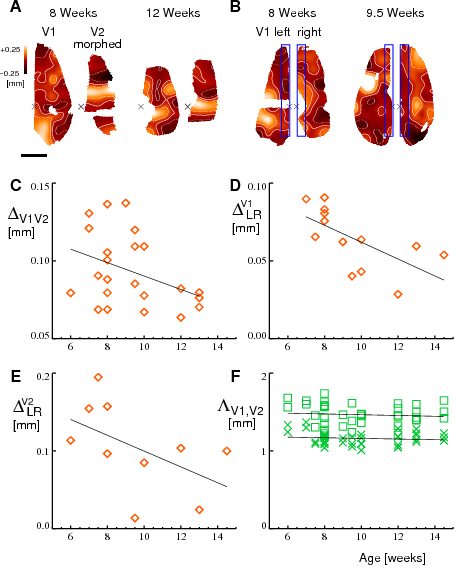}
\vspace{1cm}
\end{center}
 \caption{ }
\end{figure}

\clearpage

{\bf \Large Supplementary Material}
\vspace{1cm}

{\bf Materials and Methods.} 

First, we provide an overview over the calculation of local column
spacing and its 
decomposition into its three components analyzed in the
manuscript. In more detail, this is outlined in the remainder of
the supplement.

{\bf Decomposition of orientation column spacing.}
We analyzed the spacing of orientation columns in cat V1 and
V2 in a sample of n=41 brain hemispheres (N=27 animals) using a
recently developed 
wavelet method (S1). Because this method provides highly precise
estimates of local column spacing  with an error much smaller than the
large intrinsic variability of column spacings (SEM, 15--50$\mu$m),
differences  and   
similarities of column spacings in the sample can be identified reliably.
Orientation columns were labeled with
2-[$^{14}$C]-deoxyglucose (2-DG) (S2) and visualized in
flat-mount sections of cat visual cortical areas V1 and V2 
(Fig. S1A) (S3--S5).
The spacing of adjacent orientation columns was calculated independently at
every cortical location in each area (Fig. S1B). Thus, for each area a
two-dimensional map of local column spacing was calculated representing
the variability of local column spacings in this area.

Each map of local column spacing was decomposed into three
components   (Fig. S1C):  
(i) the mean column spacing, (ii) the systematic part of the
topographic variation of local column 
spacing, and (iii)  the individual part of
topographic variation of local column spacing.
Component (i) is a simple measure of the
variation of column spacings among different individuals and between
areas V1 and V2. As a single number, this component can be illustrated
as a flat map (the green map in 
Fig. 1SC). Component (ii) describes the systematic intra-areal
variation of column spacings and is therefore equal for all
hemispheres. It is obtained by 
superimposing and averaging the mean-corrected maps of local column
spacing from all  
41 hemispheres after aligning the V1/V2 border in all maps  (maps from the
left hemisphere were mirror-inverted). The variation of this component can be
illustrated by a color  map (the blue map in Fig. 1SC). 
Component (iii) is the map of residual variation constituting the
intra-areal variation of 
local column spacing in a map in addition to component (ii). It can be
illustrated by a color map (the orange map in Fig. 1SC).
Components (i) and (iii)  are
characteristic for an individual hemisphere in contrast to part (ii) that
characterizes the entire population of animals. 
The average of component (i) is equal to the average in the entire
population. Component (ii) and (ii) are  zero when averaged over the
map. Accordingly, the variance of local column spacings in the entire
population  
comprises three components from which the variation of mean
column spacings and the residual variation accounted for the largest
parts in V1 and V2 (Fig. S1D). The
contribution of the systematic component was relatively small in 
both areas. In order to conveniently compare topographically matching
parts in V1 and V2, maps from V2 were mirror inverted and morphed
by superimposing major landmarks such as the  
representations of the vertical meridian (located along the V1/V2
border), the central visual visual field, the 
horizontal meridian and the far periphery of V1 and V2 (Fig. S1E).

{\bf Animals.} 
20 animals (31 hemispheres) were born in the animal house of the
Max-Planck-Institut f\"ur Hirnforschung in Frankfurt am Main, Germany.
7 animals (10 hemispheres) were bought from two animal
breeding companies in Germany (Ivanovas, Gaukler). 
All animals stayed at the animal house until the 2-DG experiments. 
The visual stimuli during the 2-DG experiments were always identical
in spatial and temporal frequency, and only differed in orientation.
Mostly cardinal orientations were used. Table S1 lists details of the
dataset.

{\bf Image processing.} 
Photoprints of the 2-DG autoradiographs were digitized
using a flat-bed scanner
(OPAL ultra, Linotype-Hell AG, Eschborn, Germany, operated using Corel
Photoshop) with an effective spatial resolution of
9.45 pixels/mm cortex and 256 gray levels per pixel.
For every autoradiograph this yielded a two-dimensional (2D) array of
gray values 
$I_0({\mathbf x})$, where ${\mathbf x}$ (a 2D vector) is the position
within the area and $I_0$ its intensity of labeling.

For every autoradiograph we defined two regions of interest (ROI) 
encompassing the patterns labeled in areas V1 and V2 (S1, S3). 
The manually defined polygons encompassing the entire patterns of orientation
columns within areas V1 and V2, respectively, were stored together
with every autoradiograph. 
Only the patterns within areas V1 and V2 were used for subsequent quantitative
analysis. Regions with very low signal, and minor artifacts
(scratches, folds, and air bubbles) were excluded from further analysis.

All digitized patterns were highpass filtered 
using the Gaussian kernel 
\begin{equation}
K({\mathbf y}) = \frac{1}{2\pi
  \sigma_K^2}\exp(-{\mathbf y}^2/2\sigma_K^2)
 \end{equation}
with a spatial width of
$\sigma_K$=0.43mm for V1 and $\sigma_K$=0.57mm for V2. 
The patterns were then centered to yield $\int_{V1} d^2y \, I({\mathbf
  y})=0$. 
To remove overall variations in labelling intensity, patterns from V2 were   
 thresholded to uniform contrast by setting $I({\mathbf
  x})=1$ in regions  larger than 0, and $I({\mathbf x})=-1$ in
  regions smaller than 0.  
Finally, values in artifact regions and in regions outside of areas V1
and V2  were set to zero.

{\bf Spacing analysis.}
Patterns of orientation  columns were 
analyzed  using a wavelet method introduced recently. 
For a detailed description of the methods see S1 and S6.

For each analyzed pattern of orientation columns we determined a 
2D map representing the column spacing at each cortical location. 
We first calculated wavelet representations of a given pattern $
I({\mathbf x})$ by 
\begin{equation}
\hat{I}({\mathbf x},\theta,l) = \int_{A} d^2y \, I({\mathbf y}) \,
\psi_{{\mathbf x},\theta,l}({\mathbf y}) \,\,\,\, ,
\end{equation}
where ${\mathbf x}, \theta, l$ are the position, orientation, and scale
of the
wavelet $\psi_{{\mathbf x},\theta,l}({\mathbf y})$, 
$\hat{I}({\mathbf x},\theta,l)$ denotes the array of wavelet
coefficients and A denotes the ROI in V1 or V2.
We used complex-valued Morlet-wavelets defined by a mother-wavelet
\begin{equation}
\psi({\mathbf x})=
\exp\left(-\frac{{\mathbf x}^2}{2} \right ) \, 
e^{i{\mathbf k}_\psi \cdot {\mathbf x}} 
\label{morlet}
\end{equation}
and
\begin{equation}
\label{wavelet}
\psi_{{\mathbf x},\theta,l}({\mathbf y}) = l^{-1}
\psi\left(\Omega^{-1} (\theta) \, \frac{\mathbf y - \mathbf x}{l}\right)
\end{equation}
with the 2D rotation matrix $\Omega$.
The characteristic wavelength of a wavelet with scale $l$ is
$\Lambda_{\psi} l$ with $\Lambda_{\psi} = 2\pi / |{\mathbf k}_\psi|$.
We used wavelets with about 7 lobes, i.e. ${\mathbf
  k}_\psi=(7,0)$, to 
ensure a narrow frequency representation while keeping a good
spatial resolution of the wavelet. From these  
representations we  calculated the orientation averaged modulus
\begin{equation}
\bar{I}({\mathbf x},l) =
\int_0^{\pi} \frac{d\theta}{\pi} \, | \hat{I}({\mathbf x},\theta,l)|
\end{equation}
of the wavelet coefficients 
for every position ${\mathbf x}$, and then determined the scale
\begin{equation}
\bar{l}({\mathbf x}) = argmax\left( \bar{I}({\mathbf x},l) \right)
\end{equation}
maximizing $\bar{I}({\mathbf x},l)$. 
The corresponding characteristic wavelength
\begin{equation}
\Lambda({\mathbf x}) = \bar{l}({\mathbf x}) \, \Lambda_{\psi}
\end{equation}
was used as an estimate for the local column spacing at the position
$\mathbf x$.
For every position (spatial grid-size $0.12mm$)
wavelet coefficients for 12 orientations 
$\theta_i \in \{0, \pi/12, ..., 11\pi/12 \}$ were calculated for V1 on
15 scales $l_j$ 
(with $l_i\Lambda_{\psi}$ equally spaced in $[0.5mm,2mm]$)  and for V2
on 21 scales $l_j$ (spaced in $[0.5mm,2.5mm]$).
The scale maximizing
$\bar{I}({\mathbf x},l)$ was then estimated as the maximum of a
polynomial in $l$ fitting the $\bar{I}({\mathbf x},l_j)$  for
a given position ${\mathbf x}$ (least square fit).
The local column spacing was calculated for typically 4 flatmount sections
in each hemisphere. Values at corresponding locations in different sections
were  averaged and combined resulting in a single map of local column spacing
$\Lambda({\mathbf x})$  for V1 and V2 in each brain hemisphere. 
Locations sampled by $<$2 sections were excluded from further analysis. After
superposition, the  local column spacing $\Lambda ({\mathbf x})$ 
was smoothed using a Gaussian kernel with $\sigma$=1.25mm. 

For every map of local column spacing $\Lambda({\mathbf x})$, the
mean column spacing $\Lambda= \left< \Lambda({\mathbf x}) 
\right>_{\mathbf x} $ was calculated. It measures whether 
 a pattern predominantly contains large or small orientation columns.  
The map of the systematic topographic component of local column spacing
$\Lambda_{sys} 
({\mathbf x})$ was obtained by 
$\Lambda_{sys}
({\mathbf x}) = \left< \Lambda({\mathbf x}) - \Lambda   \right>_{hemis} $, 
i.e. by subtracting from each map of local column spacing $\Lambda({\mathbf
  x})$  its mean value $\Lambda$ and then superimposing and averaging  
over different hemispheres.  For superposition, we localized the
representations of the vertical meridians (VM)
and the areae centrales on the autoradiographs and aligned
the 2D maps of local column spacing from different animals using these
landmarks (S6, S3) Maps
from right hemispheres were mirror inverted. The alignment of  
spacing maps based on these landmarks  
matches corresponding locations from different hemispheres.
The systematic topographic component of local column spacing
$\Lambda_{sys} ({\mathbf x})$ was calculated only at locations
$\mathbf x$ where at least 8 hemispheres contributed.
Maps of individual topographic component of local column spacing,
$\Lambda_{indv} ({\mathbf x})$,  
were obtained by $\Lambda_{indv} ({\mathbf x}) = \Lambda
({\mathbf   x})- \Lambda - \Lambda_{sys} ({\mathbf x})$, i.e. by subtracting
from each map of local column spacing its mean column spacing $\Lambda$ and
the map  of systematic topographic component $\Lambda_{sys} ({\mathbf x})$.

{\bf Morphing.}
Column spacing maps from V2 were morphed on those from V1 by
thin-plate spline interpolation (S7). By this method, defined reference points
in V2 were morphed on corresponding points in V1, and the remaining
locations are morphed such that the distortion of the morphed map is
minimal. 
We used 30 reference points in  
areas V1 and V2 
distributed along the common V1/V2 border, and
along the lateral border of V2 and the medial border of V1.
The same morphing was used for all V1/V2 pairs. 
This provides only a rough mapping of corresponding locations in
individual V1/V2 pairs (see e.g. the pronounced size variation of V1
(S1)). No attempt was made to optimize the similarity of spacing maps
of V1/V2 pairs.

{\bf Accuracy and measurement errors.} 
All quantities presented are subject to measurement errors. 
The estimation of measurement errors was carried out following
(S6). 
The error of the local column spacing, $\Delta
\Lambda $, and the error of the mean column spacing, $\Delta
\Lambda ({\mathbf x})$, were estimated based on the multiple flatmount sections
analyzed for every hemisphere. Spacing values were calculated for every
section individually and SEM were estimated from the values for different
sections. 
SEMs for mean column spacings $\Lambda$ were 15$\mu$m for V1 and
35$\mu$m for V2. Errors were larger in V2 due to its smaller size and
the weaker labeling.  
Errors $\Delta \Lambda({\mathbf x})$ of the local column spacing  
were on average 58$\mu$m in V1 and 64$\mu$m in V2. 
The error of the systematic topographic component $\Lambda_{sys} 
({\mathbf x})$ was calculated by error propagation from the error of
the local  column spacing $\Lambda({\mathbf x})$ , that is  
$\Delta \Lambda_{sys} ({\mathbf x})= 
\sqrt{ \left \langle \Delta \Lambda ({\mathbf x}) ^2 \right
   \rangle_{hemis}} \, / \sqrt{ N_{{hemis}}}$, 
where the average is taken over the population of the $N_{{hemis}}$
hemispheres contributing to $\Lambda_{sys}
({\mathbf x})$. 
Its error was 
relatively small (SEM, 19$\mu$m for V1, and
23$\mu$m for V2). 
The maps of individual topographic component were mainly inflicted by
the error of local 
column spacing and the systematic topographic column spacing.

{\bf Decomposition of variance.}
The variances of all spacing parameters (e.g $\Lambda({\mathbf x}),
\Lambda, \Lambda_{{sys}} ({\mathbf x})$) were error  
 corrected following (S6). 
The variance $v_{mean}$ of the mean column
spacing $\Lambda$  was
calculated by  $v_{mean} \approx s^2_{mean} - \left \langle \Delta
 \Lambda^2 \right \rangle_{hemis} $, where 
 $s_{mean}$ is the SD of the values of the  mean column spacing
$\Lambda$ for different animals and $\left \langle \Delta
 \Lambda^2 \right \rangle_{hemis}$ is the squared error of
   $\Lambda$ averaged over hemispheres from all animals. 
For its square root $\sqrt{\left \langle \Delta  \Lambda^2 \right
   \rangle_{hemis}}$ we obtained 0.018mm for V1 and 0.046mm for V2.
The variance $v_{sys}$ of the systematic intraareal variability
of local column spacing was 
calculated from the SD $s_{sys}$ of the
systematic topographic component  $\Lambda_{sys} ({\mathbf x})$ and
the its error   
$ \Delta \Lambda_{{sys}} ({\mathbf x})$ by
$v_{sys} \approx s^2_{sys} - \left
  \langle \Delta \Lambda_{{sys}} ({\mathbf x})^2  \right
  \rangle_{\mathbf 
  x}$. The square root of the spatially averaged
squared error, $\sqrt{  \left  \langle \Delta
  \Lambda_{{sys}} ({\mathbf x})^2 \right \rangle_{\mathbf 
  x} }$, yielded 0.020mm for V1 and 0.024mm for V2, respectively. 
The variance  $v_{all}$ of 
all orientation column spacings in all hemispheres (from V1 or
from V2,
respectively) is given by  
$ v_{all} =  s^2_{all} -  \left \langle
    \Delta \Lambda ({\mathbf x})^2 \right \rangle_{{all}}$,
where  $\sqrt{\left \langle \Delta \Lambda ({\mathbf x})^2 \right
\rangle_{{all}}}$  is the square root of the error of
the local spacing squared and averaged over all locations in all
hemispheres. For V1 we obtained 0.088mm, for V2 0.093mm.  
Denoted by  $s_{all}$ is the SD of local spacing values
$\Lambda ({\mathbf x})$ from all hemispheres. 

The total variance  
$v_{all} \approx 
v_{mean} +
v_{sys} +
v_{indv}$
is composed of the variance of the mean column spacing $v_{mean}$,
the variance of the systematic topographic component of column spacing $v_{sys}$, and the average variance of 
the individual topographic component of column spacings $v_{indv}$. 
This decomposition provides an estimate for the relative
magnitudes of the different contributions to the total variance in the  
 population of column spacing maps from V1 or V2 (Fig. 1D).

{\bf Permutation tests.} 
Permutation tests were used to test for statistical
significance. In these tests the value of a statistic (e.g. for
cross-correlation or 
for an average differences) was compared to values obtained 
for randomized data. Usually, a  distribution of 10$^4$ random
realizations  was sampled. The significance value is given by the probability
of obtaining the   
real value (or a value more extreme) by chance.   
The significance value for the correlation between mean 
column spacings $\Lambda$ in V1 and V2 was calculated by permuting all
mean column spacings from V2 and is given by the fraction of
correlation coefficients found to 
be larger than the real value. 
The significance of the distance $\Delta$ between the map of the 
residual topographic component from V1 and and the morphed map from V2
was calculated by 
permuting among all maps from V2. The average distance $\Delta$ was
calculated from all V1/V2 pairs with a common area of at least 70mm$^2$
(in the coordinate system of V1) 
and compared to averages obtained in 10$^4$ comparable groups of   
pseudo V1/V2 pairs. The significance value is given by the
fraction of averages smaller than the average of the real distance $\Delta$.
Distances $\Delta$ between  individual topographic variations in 
the left and right hemispheres were compared to pseudo 
left/right pairs generated from all hemispheres.  
All significance tests regarding the distance $\Delta$ were based on aged
matched randomizations. Cases 9 weeks old or younger (n=19) were 
exchanged by pseudo pairs generated from this group only. Random pairs older
than 9 weeks were generated only from the cases older than 9
weeks (n=22).

\newpage

{\bf References}\\

\vspace{0.4cm}S1.
M.~Kaschube, F.~Wolf, T.~Geisel, S.~L\"owel, {\it J. Neurosci.\/} {\bf 22},
  7206 (2002).

\vspace{0.4cm}S2.
S.~L\"owel, {\it The Cat Primary Visual Cortex\/}, B.R.~Payne and
A.~Peters, eds. 

\hspace{0.7cm}(Academic Press, San Diego, 2002), chap. 2, pp. 167--193

\vspace{0.4cm}S3.
S.~L\"owel, B.~Freeman, B.~Singer, {\it J. Comp. Neurol.\/} {\bf 255}, 401
  (1987).

\vspace{0.4cm}S4.
S.~L\"owel, H.-J. Bischof, B.~Leutenecker, W.~Singer, {\it Exp. Brain Res.\/}
  {\bf 71}, 33 (1988).

\vspace{0.4cm}S5.
S.~L\"owel, W.~Singer, {\it Dev. Brain Res.\/} {\bf 56}, 99 (1990).

\vspace{0.4cm}S6.
M.~Kaschube, {\it et~al.\/}, {\it Eur. J. Neurosci.\/} {\bf 18}, 3251 (2003).

\vspace{0.4cm}S7.
F.~Perrin, O.~Bertrand, J.~Pernier, {\it IEEE Trans. Biol. Eng.\/}
{\bf 34}, 283, (1987).\\

 \newpage
{\bf Table S1.}
Dataset used for quantitative analysis of orientation maps in cat
areas V1 and V2. We
analyzed 41 hemispheres from 27 animals (C1-C27). 
For every animal, the table lists the hemispheres (left=le, right=ri,
or both=le+ri), the number of 2-DG autoradiographs analyzed for V1 and
V2, the orientation of the visual 
stimulus ($0^o=$horizontal, $90^o=$vertical, $45^o=$right oblique,
$135^o$=left oblique), and the age at the time of the experiment.
Cats C1-C20 were born and raised in the animal
house of the Max-Planck-Institut f\"ur Hirnforschung in Frankfurt am
Main, Germany. 
Animals C21-C27 were bought from the animal breeding
companies Ivanovas 
(C23 and C25) and Gaukler (C26 and 27), both from Germany.\\

{\bf Fig. S1.} 
Decomposition of orientation column spacing in cat visual
cortex. ({\bf A}) Overall
layout of
2-[$^{14}$C]-deoxyglucose labeled (dark gray) orientation
columns in flat mount sections of areas V1 (right) and V2
(left). Black and white
arrow heads indicate the external border of V1 and V2, respectively.
Cortical representations of the vertical meridian (VM)
(i.e. the V1/V2 border) and the horizontal meridian (HM) of the visual field
are represented by the white dashed lines (a=anterior, m=medial; scale
bar, 10mm).
({\bf B}) 2D-maps of local column spacing in areas V2
and V1 (gray scale coded).  Contour lines
are drawn at the mean spacing (thick white line) and mean $\pm$ SD (thin
white lines).
Black crosses mark the central visual field representation.
({\bf C}) Each map of local column spacing is composed of (i)
the mean column spacing (green)
(ii) the systematic part of the topographic component of local column
spacing (population averaged, blue), and (iii)  the individual part of
topographic  component (orange) [here illustrated for the V1 map in {\bf B}].
({\bf D}) According to {\bf C} the variance of all column spacings in the
population is the sum of (i)
the variance of the mean column spacings of the different areas (green),
(ii) the variance of the
systematic topographic component (blue), and (iii) the
average variance of the individual topographic component (orange).
The percentages of these variance components are represented by
colored bars for
V2 (i, 38\%; ii, 8\%; iii, 54\%) and V1 (i, 34\%; ii, 14\%; iii, 52\%), and for
ocular  dominance  columns in cat V1  (i, 18\%; ii, 24\%; iii, 58\%).
({\bf E}) For comparing layouts in V1 and V2, V2 spacing maps were mirror
inverted and morphed (shown schematically) aligning regions
representing similar parts of the visual field  in areas V1 and V2.

{\bf Fig. S2.} 
({\bf A}-{\bf F}) The pattern of individual topographic component of column
spacing for  
V2, V1, and the morphed version of V2 for six hemispheres. 
(color scale, black cross and contour lines as in Fig. 1).  Maps from
right hemispheres were mirror-inverted. Scale bar, 10mm.
Note the same general pattern in both areas V1 and V2 with  maxima (white) and minima
(dark orange)  often at  corresponding retinotopic locations.
Note also that the pattern of individual component differs strongly in
different individuals.

\newpage
\begin{center}
\begin{tabular}{|c|c|c|c|c|}
\hline
animal & hemispheres & autorad. & stimulus & age [weeks]\\
\hline
\hline
C1 & le+ri & 4/4 & $0^o$ &  9.5 \\ \hline
C2 & le+ri & 5/2 & $0^o$ &  10 \\ \hline
C3 & le+ri & 4/4 & $0^o$ &  8  \\ \hline
C4 & le+ri & 3/4 & $0^o$ &  7.5 \\ \hline
C5 & le+ri & 5/6 & $90^o$ &  8 \\ \hline
C6 & ri & 5/5 & $90^o$ &  8  \\ \hline
C7 & le+ri & 4/4 & $0^o$ &  8 \\ \hline 
C8 & le+ri & 4/4 & $90^o$ &  14.5 \\ \hline
C9 & ri & 4 & $90^o$ &  14.5 \\ \hline
C10 & ri & 5 & $0^o$ & 12 \\ \hline
C11 & ri & 4 & $0^o$ & 12 \\ \hline 
C12 & ri & 4 & $90^o$ &  13 \\ \hline
C13 & ri & 3 & $90^o$ &  13 \\ \hline   
C14 & le  & 4 & $90^o$ &  10 \\ \hline
C15 & le+ri & 4/4 & $90^o$ &  8 \\ \hline 
C16 & le+ri & 4/4 & $90^o$ &  12 \\ \hline
C17 & le+ri & 4/4 & $0^o$ &  6  \\ \hline 
C18 & le+ri & 5/5 & $45^o$ &  7  \\ \hline
C19 & le & 4 & $0^o$ &  9.5  \\ \hline  
C20 & ri & 4 & $0^o$ &  8  \\ \hline    
C21 & le+ri & 4/3 & $135^o/0^o$ &  13 \\ \hline s3
C22 & le & 4/4 & $135^o/0^o$ &  8 \\ \hline 

C23 & le+ri & 4/4 & $0^o$ &  10  \\ \hline
C24 & le+ri & 4/4 & $0^o$ &  9 \\ \hline 
C25 & le & 4 & $0^o$ &  13  \\ \hline 

C26 & ri & 5 & $90^o$ &  12 \\ \hline 
C27 & ri & 3 & $90^o$ &  12 \\ \hline
\end{tabular}

\vspace{2cm}

Table S1.
\end{center}

\newpage
\begin{figure}
\begin{center}
\includegraphics[bb=0 0 754 400,width=14cm]{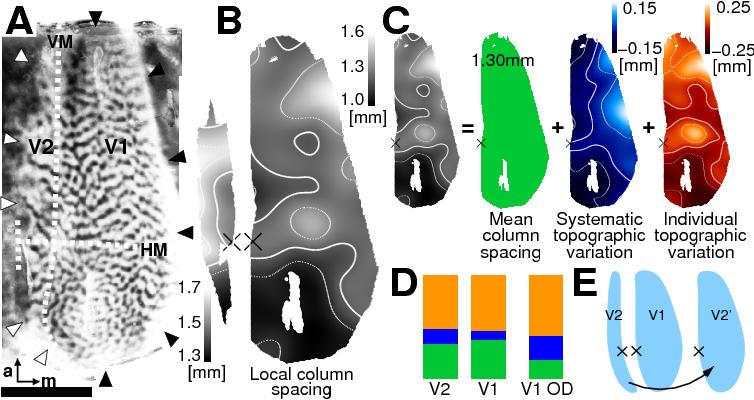}\\

\vspace{2cm}Fig. S1.
\end{center}
\end{figure}

\newpage
\begin{figure}
\begin{center}
\includegraphics[bb=0 0 361 388,width=12cm]{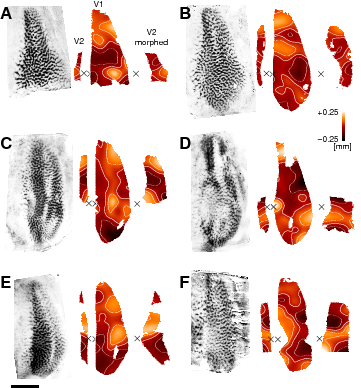}\\

\vspace{2cm}Fig. S2.
\end{center}
\end{figure}

\end{document}